**Recursive polarization of nuclear spins in diamond at arbitrary magnetic fields**

*Daniela Pagliero[1], Abdelghani Laraoui[1], Jacob D. Henshaw[1], Carlos A. Meriles[1]*

[1]Dept. of Physics, CUNY-City College of New York, New York, NY 10031, USA.

**Abstract**

We introduce an alternate route to dynamically polarize the nuclear spin host of nitrogen-vacancy (NV) centers in diamond. Our approach articulates optical, microwave and radio-frequency pulses to recursively transfer spin polarization from the NV electronic spin. Using two complementary variants of the same underlying principle, we demonstrate nitrogen nuclear spin initialization approaching 80% at room temperature both in ensemble and single NV centers. Unlike existing schemes, our approach does not rely on level anti-crossings and is thus applicable at arbitrary magnetic fields. This versatility should prove useful in applications ranging from nanoscale metrology to sensitivity-enhanced NMR.

Formed by a nitrogen impurity adjacent to a vacant site, the nitrogen-vacancy (NV) center in diamond is emerging as a promising platform for multiple applications in photonics, quantum information science, and nanoscale sensing [1]. A fortuitous combination of electronic structure, intersystem crossing rates, and selection rules allows the NV ground state spin-triplet ($S = 1$) to completely convert into the $m_S = 0$ magnetic sublevel upon optical illumination. This easily obtainable pure quantum state provides the basis to initialize the NV spin and, perhaps more importantly, other neighboring spins that cannot be polarized by optical means. For example, it has been shown that the nuclear spin of the nitrogen host polarizes almost completely near 50 mT, where the NV experiences a level anti-crossing (LAC) in the excited state[2]. Further, nuclear spin hyperpolarization has been observed in strongly coupled carbon spins at 50 mT and 100 mT, mediated by NV LACs in the excited[2,3] and ground[4] states, respectively. Alternatively, efficient polarization transfer from the NV was demonstrated via the use of Hartman-Hahn (HH) protocols adapted to the doubly rotating frame — to polarize other paramagnetic species[5,6] — or to the mixed rotating/lab frame — to selectively polarize adjacent carbons[7]. Other forms of polarization transfer to nuclear spins from NVs have been investigated as well[8].

Here we introduce an alternate strategy that simultaneously exploits the singular energy level structure of an NV-$^{14}$N pair in the ground state as well as the robustness of the nuclear spin state against optical excitation of the NV. We build on this latter notion to articulate two complementary schemes: The first one — which we will refer to as 'spin exchange' — concatenates optical excitation with selective microwave (mw) and radio-frequency (rf) pulses to exchange the initial NV and nitrogen spin states. The second one — here denoted as 'population trapping' — uses a similar pulse sequence to sequentially drag the NV-$^{14}$N system into a target polarized state. Interestingly, this class of schemes promises some advantages when compared to LAC- and HH-based methods, particularly when nuclear polarization at high magnetic field is required.

The experiments herein are carried out using a confocal microscope adapted for optically-detected magnetic resonance, as described previously[9]. Briefly, we manipulate the NV (nitrogen) spin via the mw (rf) field created by a copper wire stretching the diamond surface. We apply short pulses (<1 μs) of green laser light (532 nm) to either polarize or readout the NV spin state (the NV fluorescence is brighter in the $m_S = 0$ state of the ground state triplet[10]). We conduct both single-NV and ensemble measurements using two [111] diamond crystals (type IIa and Ib, respectively) with typical NV spin coherence lifetimes of ~300 μs.

To more clearly introduce our approach to polarization transfer we start by considering the energy level structure of an NV-$^{14}$N spin pair (Fig. 1a). In the ground state, the NV exhibits a spin triplet with zero field splitting Δ~2.87 GHz. Each of these levels, however, unfolds into three different states depending on the energy of the $^{14}$N nuclear spin (itself a spin $I$=1 system subject to a quadrupolar coupling $Q$~5 MHz[11]). The NV-$^{14}$N hyperfine coupling — of magnitude $A$~2.2 MHz — vanishes in the $m_S = 0$ state of the NV but it must be accounted for in the $m_S = \pm 1$ levels. Further, the presence of a magnetic field aligned with the NV axis breaks the degeneracy between the $m_S = \pm 1$ levels (as well as the $m_I = \pm 1$ states in the $m_S = 0$ subspace) thus making it possible to distinguish all transitions of the spin pair (Fig. 1a).

Our first approach to polarization transfer rests on the one-to-one correspondence between the NV and $^{14}$N spin states. A schematics of the pulse protocol is presented in Fig. 1b: After optical initialization of the NV into $m_S = 0$, we apply a set of four consecutive mw and rf inversion pulses each of which is tailored to act *selectively* on a given target transition, as shown in the graph. Assuming an initial mixed state of the form $\alpha|m_S = 0, m_I = 1\rangle + \beta|m_S = 0, m_I = 0\rangle + \gamma|m_S = 0, m_I = -1\rangle$, the first two pulses lead to $\alpha|m_S = 1, m_I = 0\rangle + \beta|m_S = 0, m_I = 0\rangle + \gamma|m_S = 0, m_I = -1\rangle$, while the second pair produces the final target state $\alpha|m_S = 1, m_I = 0\rangle + \beta|m_S = 0, m_I = 0\rangle + \gamma|m_S = -1, m_I = 0\rangle$. Therefore, the pulse sequence acts as a state exchange (SE) protocol transferring the initial spin polarization to the nitrogen and leaving the NV spin in a mixed, unpolarized state.

To monitor the spin transfer efficiency we apply a pulsed ESR protocol in which the NV is repolarized and its fluorescence measured immediately after the application of a narrow band inversion pulse. Fig. 1c shows the resulting hyperfine-split spectrum as we scan the frequency of the probe pulse throughout the $|m_S = 0\rangle \leftrightarrow |m_S = -1\rangle$ group of transitions. Unlike the case with no spin transfer, the dip corresponding to the transition $|m_S = 0, m_I = 0\rangle \leftrightarrow |m_S = -1, m_I = 0\rangle$ becomes prominent, implying that the nitrogen spin has been successfully initialized into the $m_I = 0$ state. More quantitatively, from the ratio between the central peak and satellite amplitudes, we estimate the fractional nuclear spin population in the $m_I = 0$ state of order 75%, corresponding to a sub-milli-Kelvin nuclear spin temperature.

Implicit in the experiment above is the condition that the NV can be re-pumped into the $m_S = 0$ state without affecting the nuclear spin state. This notion — already used to demonstrate repetitive readout at the single NV level[12,13] — can be further exploited here to implement the alternate nuclear spin pumping protocol of Fig. 2. As before, a total of four mw and rf pulses is used, but in this case all microwave pulses act exclusively on one of the two possible NV transitions, $|m_S = 0\rangle \leftrightarrow |m_S = \pm 1\rangle$. This scheme can be understood as a 'population trapping' (PT) scheme attained via consecutive CNOT gates in the form of selective π-pulses: On the condition that the initial nuclear spin quantum number is $m_I = +1$, the first pair of mw and rf pulses — here assumed to act on the $|m_S = 0\rangle \leftrightarrow |m_S = -1\rangle$ set of transitions — maps the state $|m_S = 0, m_I = +1\rangle$ into $|m_S = -1, m_I = 0\rangle$. The latter is subsequently transformed into $|m_S = 0, m_I = 0\rangle$ via the application of a pump laser pulse (denoted as $p1$ in Fig. 2b). Similarly, if the initial nuclear spin state is $m_I = -1$, the second train of mw, rf, and light pulses maps the state $|m_S = 0, m_I = -1\rangle$ into $|m_S = 0, m_I = 0\rangle$, which consequently becomes the only possible final state of the spin pair.

Fig. 2c shows the hyperfine-split spectra corresponding to the NV $|m_S = 0\rangle \leftrightarrow |m_S = -1\rangle$ transition after applying the PT protocol at three different magnetic fields. Similar to Fig. 1c, we attain nitrogen initialization ranging from 75 to 80 percent, depending on the applied magnetic field. From among the three cases we investigated, the highest pumping efficiency is attained at 5 mT, where level mixing between the electronic and nuclear spin states in the optically excited NV triplet is comparatively lower (see below). Similar initialization efficiency is attained in single NVs (cases (i) and (iii)) and NV ensembles (case (ii)) indicating that this protocol can be used to generate bulk nuclear spin polarization[4,3]. Naturally, the pulse sequence can be altered to polarize a different target nuclear spin state. This is shown in case (iii) where the second microwave pulse is tuned into the $|m_S = 0, m_I = 0\rangle \leftrightarrow |m_S = -1, m_I = 0\rangle$ transition, thus trapping the nitrogen nuclear spin population into the $m_I = -1$ state. Note that efficient nitrogen initialization is attained even in the presence of partial level mixing in the NV excited triplet state[2] (the mechanism responsible for the preferential pumping of the $^{14}$N into $m_I = +1$ observed in the absence of rf, blue trace in case (iii)).

Interestingly, the ability to repump the NV into $|m_S = 0\rangle$ without (significantly) affecting the nitrogen state makes both spin transfer schemes above one-directional, meaning that they can be applied recursively so as to mitigate system imperfections (e.g., mw and/or rf inhomoneity, or frequency offsets). We demonstrate this notion in Fig. 3 where we repeat the PT sequence a variable number of times while purposely setting the duration of the rf pulses away from the ideal value. To probe the $^{14}$N populations we apply in this case a Ramsey sequence, and subsequently Fourier-transform the NV response. Given the long-term memory of the nuclear magnetization and the need to average the NV signal over multiple runs of the pulse sequence, we prolong the readout pulse $pr$ — typically 250 ns long — over a longer time interval (10 µs). The goal is to erase at the end of each run the polarization gain attained during the spin transfer (see below) so as to reset the NV-$^{14}$N system to the same (known) initial state.

The results are shown in Fig. 3b where we plot the observed NV spectra for a different number $n$ of transfer cycles. In this particular example, we purposely set the rf pulses to $\beta = \pi/2$ (i.e., half the ideal value) while keeping the mw duration near its optimum (i.e., $\alpha = \pi$). By comparing the relative amplitudes in the hyperfine-split spectra, we find that a one-time application of the spin transfer protocol produces only a mild change in the $^{14}$N populations ($n = 1$ in Fig. 3b). As $n$ increases, however, we observe a progressive growth of the $m_I = 0$ peak towards a maximum, which remains unchanged when $n \geq 6$. This maximum is slightly smaller than that attained when $\beta = \pi$ (denoted as 'Optimum transfer'), a trend we confirmed through additional observations using various values of $\alpha$ and $\beta$. The exact asymptotic limit — as well as the convergence rate — is found to be a function of the chosen conditions, which points to a complex interplay between the number of pumping cycles, the mw/rf offset, an the effect of laser illumination (itself a function of the pulse duration and applied magnetic field).

We gain a semi-quantitative understanding of the observed response by considering the simplified model of Fig. 4a: Here a hypothetical system of coupled electron and nuclear spins — both sharing the same spin number $S' = I' = 1/2$ — undergo repeated cycles of dynamic polarization into the target state $|m'_S = +1/2, m'_I = +1/2\rangle$. Under the combined action of the mw and rf pulses (not necessarily performing ideally), the nuclear spin experiences a flip with probability $p_a \lesssim 1$. Similar to the NV, we assume that the light pulse deterministically maps the electron spin into the group of sublevels within $\{m'_S = +1/2\}$, and that this process may be accompanied by a nuclear spin flip with probability $p_b \gtrsim 0$. Under these conditions, the relative populations in the $m'_S = +1/2$ subspace at the end of the $n$-th pumping cycle are given by $P_-^{(n)} = P_-^{(n-1)} q + p_b$ and $P_+^{(n)} = 1 - P_-^{(n)}$, where $q \equiv (1 - p_a)(1 - 2p_b)$, and the $\pm$ subscripts correspond to the $|m'_I = \pm 1/2\rangle$ states. After $N$ cycles, the probability of having a nuclear spin in the depleted state $|m'_S = +1/2, m'_I = -1/2\rangle$ amounts to

$$P_-^{(N)} = P_-^{(0)} q^N + p_b (1 - q^N)/(1 - q) , \qquad (1)$$

where $P_-^{(0)}$ denotes the initial nuclear spin population in $|m'_S = +1/2, m'_I = -1/2\rangle$. Eq. (1) indicates that convergence to the optimum nuclear spin polarization in a single cycle is possible when $q = 0$ (corresponding to $p_a = 1$ if we assume $p_b < 1/2$). Note that as $N$ increases, $P_+^{(N)} = 1 - P_-^{(N)}$ asymptotically approaches the limit value

$$P_+^{lim} = 1 - p_b/(p_a + 2p_b(1 - p_a)) \leq 1 - p_b \,, \qquad (2)$$

implying that complete nuclear spin initialization can be attained for an arbitrary value of $p_a$ when the probability of an optically induced flip is negligible. In the more general case where $p_b \neq 0$, the limit nuclear polarization grows with $p_a$ and reaches a maximum when $p_a = 1$.

The conclusions above are in qualitative agreement with the observations of Fig. 4b where we plot $P_0$ — the fractional $^{14}$N population in $m_I = 0$ — as a function of the number of cycles $N$ in the population trapping sequence (see Fig. 3a). A comparison between our observations and the model prediction — here included only as a reference — yields $p_b \sim 0.20$ (see top graph in Fig. 4b) for the present conditions of illumination and magnetic field. We warn that this value must be understood as a crude estimate given the more complex level structure of the NV-$^{14}$N system.

We gain a more direct appreciation of the effect of the optical pulses on the $^{14}$N populations via the experiment of Fig. 4c. Here we measure the NV response after application of the polarization trapping sequence as we increase the duration of the light pulse $p2$. In this example, all mw and rf pulses are set to their ideal durations, which allows us to attain optimum $^{14}$N initialization (~77%) for the shortest $p2$ time. However, the spin pumping efficacy gradually decays as $p2$ becomes longer to finally vanish when the illumination interval exceeds ~5 µs. We find similar results when the rf duration is less than optimal, with an overall scaling that depends on $p_a$, as summarized in Fig. 4d.

In the context of the model above, we interpret this behavior as a progressive growth of $p_b$, which impacts the steady state nuclear spin populations according to Eq. (2).

In summary, we introduced an alternate route to nuclear spin polarization in diamond that leverages on the relative robustness of the nuclear spin against NV optical pumping. One embodiment of this concept uses a spin swap protocol to exchange the spin states of the NV and its nuclear neighbor, whereas the other iteratively drags the nuclear spin population into a final target state. Both schemes are most effective at magnetic fields removed from the ground- or excited-states NV level anti-crossings, which makes them complementary to known polarization methods relying on nuclear/electron state mixing. In particular, since nuclear spins become increasingly decoupled from the NV at fields greater than ~150 mT, these protocols promise to be useful in situations where low-field work is inconvenient or undesired. Examples are the implementation of repetitive redout schemes (when detection sensitivity is insufficient to initialize the $^{14}$N in a 'single-shot') and NMR experiments relying on inductive nuclear spin detection (impractical below ~200 mT). Both schemes can be applied recursively so as to circumvent limitations arising from imperfections in the mw or rf pulses, an advantageous feature when polarizing nuclear spin ensembles using macroscopic rf coils or mw resonators. The best performance is anticipated at magnetic fields comparable to or greater than 200 mT, where the probability of a light-induced $^{14}$N flip becomes negligible[13].

Several routes are conceivable to extend the transfer of spin polarization to carbon nuclei, either to selectively initialize target ancilla spins or the full $^{13}$C bath. For example, since the contact coupling of the first shell carbons in the NV ground state amounts to $A_C \sim 200$ MHz (two orders of magnitude larger than nitrogen), strongly coupled carbons can be readily polarized by adapting, e.g., the PT protocol to a spin-1/2 nucleus. On the other hand, *ensemble* carbon polarization can be attained by transferring the $^{14}$N spin order via a Hartman-Hahn protocol. Note that although direct Hartman-Hahn transfer from the NV to carbons has already been demonstrated[7], a two-step process involving the $^{14}$N spin may prove advantageous at high field, where matching the mw amplitude to the carbon Zeeman frequency is impractical. Along these lines, direct flip-flops between the $^{14}$N and $^{13}$C nuclei are possible near 500 mT — where the carbon Larmor frequency approximately matches the $^{14}$N quadrupolar splitting — thus allowing for spin transfer to the carbon bath without the need for extra rf pulses. This strategy could prove useful in the polarization of near-surface $^{13}$C ensembles, which, in turn, could be exploited to polarize overlaid films[14,15] or fluids brought into contact with a diamond crystal[8,16].

The authors acknowledge support from the National Science Foundation through grants NSF-1314205 and NSF-1309640.

Pagliero et al., Fig. 1

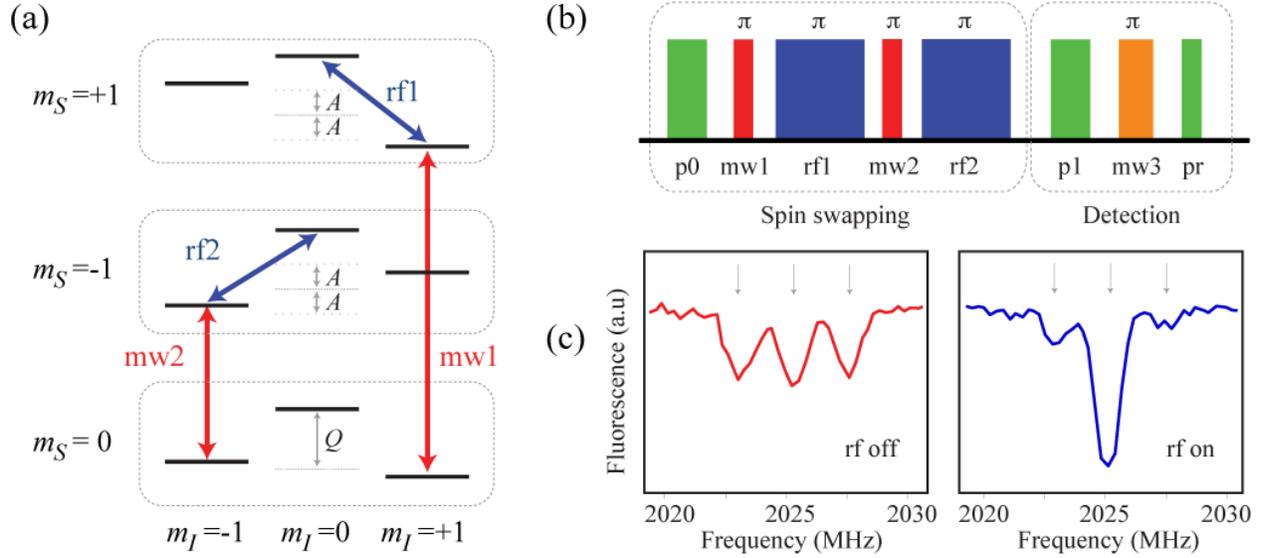

**Fig. 1:** (a) Energy level diagram of the NV-$^{14}$N system in the NV ground state; $Q$ and $A$ respectively denote the amplitudes of the nitrogen quadrupolar and hyperfine couplings. Blue (red) arrows indicate rf (mw) pulses. (b) Spin exchange protocol; *p0* and *p1* denote light pulses, *pr* indicates a readout laser pulse, and *mw3* is a narrowband mw pulse of variable frequency. (c) NV ODMR spectrum at 30 mT ($|m_S = 0\rangle \leftrightarrow |m_S = -1\rangle$ transition) of a type Ib diamond (ensemble measurement); the faint grey arrows indicate the spectral positions of the $^{14}$N hyperfine-shifted dips. After application of the protocol in (b) (right plot), a prominent dip in the spectrum reveals almost full initialization of the $^{14}$N spin into $|m_I = 0\rangle$. By contrast, three equivalent dips are present when the rf amplifier is turned off (left plot).

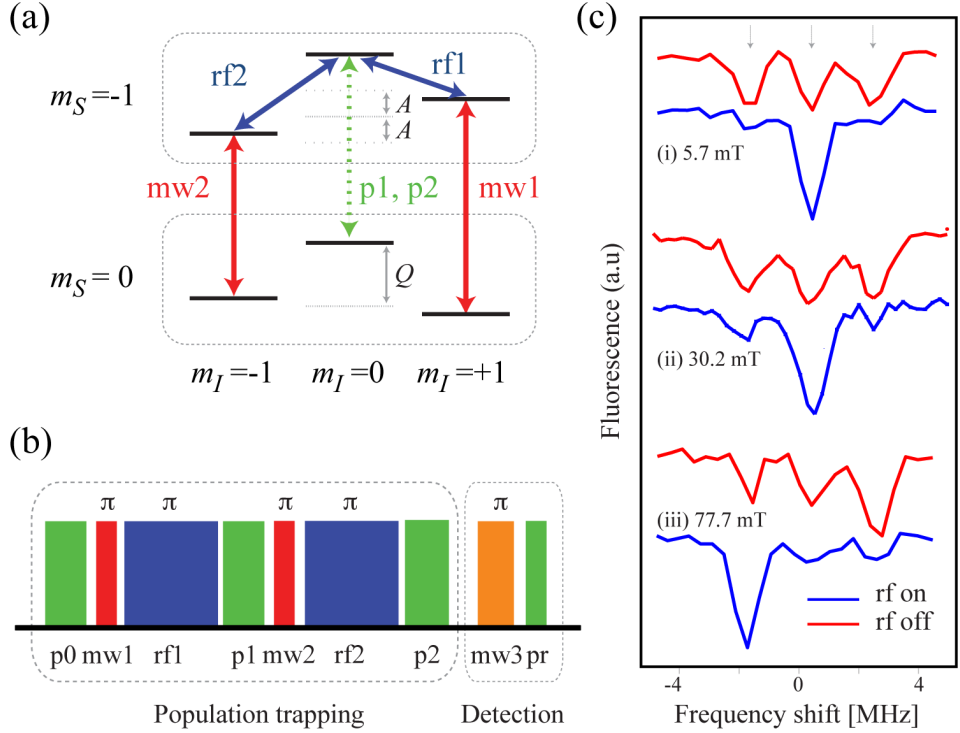

**Fig. 2:** (a) Simplified NV-$^{14}$N energy diagram. Blue (red) arrows denote rf (mw) pulses; the dashed green arrow reflects the effect of laser pulses *p1* and *p2*. (b) Pulse sequence comprising the population trapping and detection protocols; *mw3* denotes a selective mw pulse of variable frequency and *pr* is a readout light pulse. (c) NV ODMR spectra after application of the sequence in (b) at 5.7 mT (top), 30.2 mT (middle) and 77.7 mT (bottom) referred to as cases (i), (ii) and (iii), respectively. As in Fig. 1, blue and red traces indicate spectra obtained with the rf transmitter switched to on or off, respectively. Cases (i) and (iii) correspond to single NV measurements in a type IIa diamond crystal whereas case (ii) corresponds to NV ensemble measurements in a type Ib crystal. In case (iii), the pulse protocol in (a) was modified so that *mw2* acts selectively on the $|m_S = 0, m_I = 0\rangle \leftrightarrow |m_S = -1, m_I = 0\rangle$ transition.

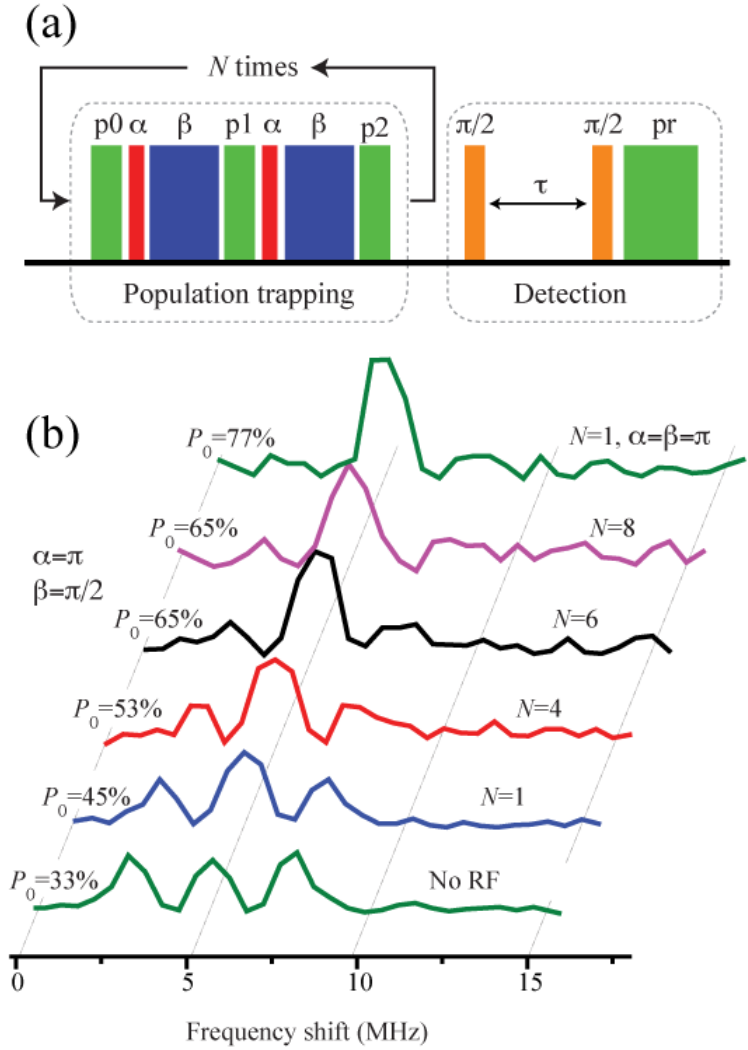

**Fig. 3:** (a) Recursive application of the population trapping protocol; red (blue) bars indicate mw (rf) pulses of duration $\alpha$ ($\beta$). Detection is carried out via a Ramsey sequence followed by a 10 μs light pulse so as to ensure complete nuclear spin depolarization between successive applications of the protocol. (b) NV spectra after Fourier transform of the signal in (a). In this case, we choose the rf duration $\beta=\pi/2$ (i.e., one half the ideal value); $N$ indicates the number of repeats in the pumping cycle and $P_0$ denotes the estimated fractional population of the $m_I = 0$ state. The spectrum corresponding to $N=1$ and $\alpha=\beta=\pi$ (optimum transfer) is also included for reference.

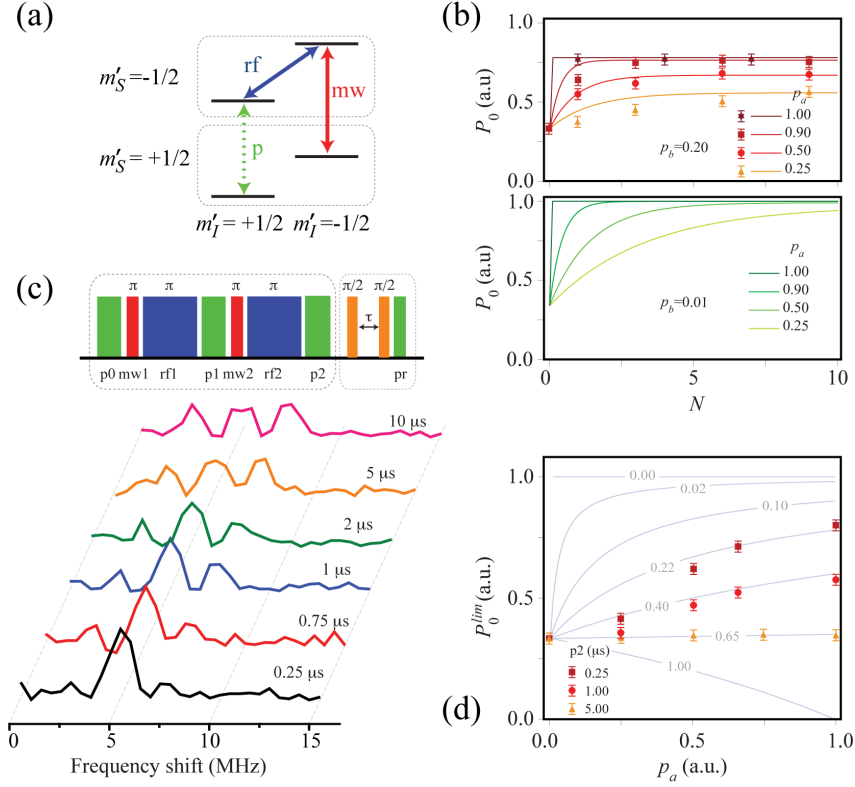

**Fig. 4:** (a) Toy-model energy diagram for a virtual, hyperfine-coupled electron/nuclear spin pair. Assuming preferential optical pumping into $\{m'_S = +1/2\}$, nuclear spin initialization into $m'_I = +1/2$ is attained via the application of selective mw, rf, and light pulses, as sketched in the figure. (b) Fractional $^{14}$N population in the in the $m_I = 0$ state — denoted as $P_0$ — after $N$ repetitions of the PT protocol. In each case the duration of the rf pulses is $\beta = p_a\pi$; symbols denote data points and lines correspond to the simplified model of (b) assuming an initial population $P_+^{(0)} = 1/3$ in state $m'_I = +1/2$. The optical flip probability $p_b$ in the upper and lower graphs is 0.20 and 0.01, respectively. (c) NV spectra for different durations of the light pulse $p2$; all other light pulses have a duration of 250 ns and all mw and rf pulses correspond to π-rotations. (d) Symbols represent the measured limit population of the $m_I = 0$ state $P_0^{lim}$ as a function of the rf-induced flip probability $p_a$ for different durations of $p2$. Faint lines correspond to $P_+^{lim}$ as determined from the model in (b); labels denote the assumed values for $p_b$.

15